\documentclass[11pt,letterpaper]{article} 
\usepackage{ol2}
\usepackage[draft]{hyperref} 
\usepackage{srcltx}

\usepackage{srcltx}

\usepackage[draft]{hyperref} 

\begin{document}

\title{Three dimensional tracking of gold nanoparticles using digital holographic microscopy}

\author{Fr\'{e}d\'{e}ric Verpillat$^a$, Fadwa Joud$^a$,  Pierre Desbiolles$^a$ and Michel Gross$^b$}

\affiliation{
$^a$ Laboratoire Kastler Brossel, Ecole Normale Sup\'{e}rieure, UPMC, 24 Rue Lhomond, 75005
Paris, France;\\
$^b$ Laboratoire Charles Coulomb, Universit\'{e} Montpellier II, Place Eug\`{e}ne Bataillon, 34095
Montpellier, France;
}

\begin{abstract}
In this paper we present a digital holographic microscope to track gold colloids in three dimensions. We report
observations of 100 nm gold particles in motion in water. The expected signal and the chosen method of reconstruction
are described. We also discuss about how to implement the numerical calculation to reach real-time
3D tracking.
\end{abstract}

\maketitle
\bigskip

\bigskip

\section{Introduction}
The tracking of nano-scale biological markers in motion is a challenging project, that could lead to a better
understanding of dynamics in biological systems. Since digital holography gives access to the whole light field in
a volume from a single hologram \cite{schnars1994direct,leith1965wavefront}, this technique is suitable to reconstruct the field scattered by gold particles
in three dimensions, and then determine their positions. This localization can be done without any mecanical
scanning of the setup. The 3D tracking in digital holography was demonstated for micron-size particles in several
publications with in-line holography \cite{cheong2010strategies,sheng2006digital,speidel2009interferometric,xu2003tracking}.
The observation of nano-scale gold particles was also demonstrated in
off-axis holography by Atlan et al.\cite{atlan2008heterodyne}, and on the membrane of a living cell by Joud et al. \cite{warnasooriya2010imaging}.  Nano particles could
be observed with the combination of off-axis geometry and phase-shifting technique \cite{zhang1998three}, that achieves almost the
theoretical noise level \cite{gross2007digital}. In this paper, we present first our digital holographic microscope, then we explain the
reconstruction method. Some results of 3D tracking of 100 nm nano-beads are presented.

\section{Experimental setup}

\begin{figure}
\begin{center}
\includegraphics[width=12 cm]{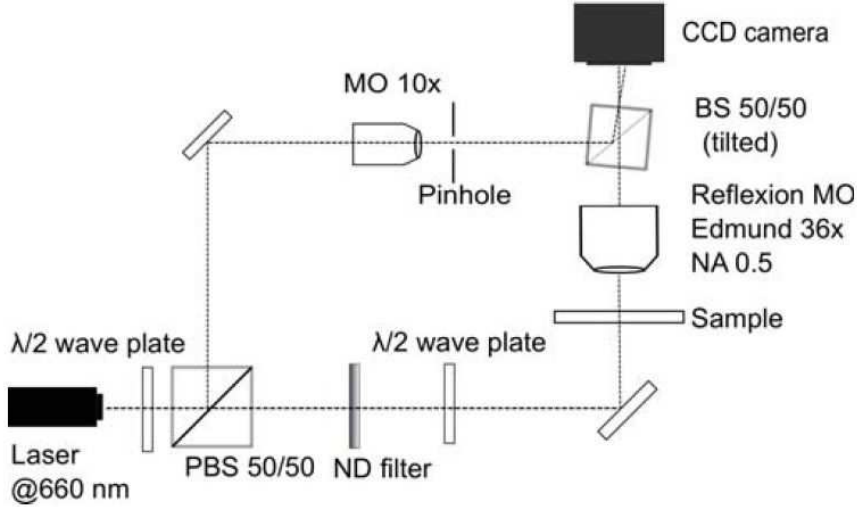}\\
  \caption{Experimental setup.}\label{Fig1}
\end{center}
\end{figure}

We chose to track 100nm gold beads in Brownian motion in water. This particles are good light scatterers
because their optical index in visible light is very low. Moreover they are non cyto-toxic and chemically inert \cite{jain2006calculated}.
The sample is prepared using a 400 $\mu$m deep chamber between two cover glasses filled with a nano-beads solution.
The experimental setup is shown on the figure 1. The optical setup use a 100 mW DPSS laser emitting at 660 nm.
The laser beam is splitted with a polarizing beam splitter. We adjust the ratio of energy between both beams
with an half-wave plate placed before the splitter. The first beam, called illumination beam, is focalised on the
sample. The second beam, called reference beam, is spatially filtered and expanded in a way to cover all the
CCD chip. A dark-field microscope objective from Edmund Optic (NA = 0.5 and 36$\times$ magnification) collects
the scattered light and blocks the illumination beam. This dark field configuration is necessary to avoid the
saturation of the CCD chip : the Rayleigh-Mie scattering model \cite{van1981light} gives a cross section of 0.015 $\mu$m$^2$ for a 100 nm
diameter gold particle at 660 nm. During 1 ms, the number of incident photons is typically 1014 photons at
660 nm, and the area of the illumination beams waist is $\simeq  4 \times  10^{-8}$ m$^2$. The number of scattered photons by
the nano-bead collected through a numerical aperture of 0.5 is equal to $8 \times 10^6$ photon during 1 ms. The ratio
of number of photons between the illumination beam and the scattered light is about $10^8$, so its not possible
to record the scattered signal on our 16 bits CCD chip without blocking the illumination beam. The dark field
objective allows to use the full dynamic of the CCD chip to record the low scattered signal. A non-polarizing
beam splitter behind the microscope objective combines the reference and the scattered signal. This splitter is
titled of a few degrees to be in off-axis geometry \cite{cuche2000spatial}. The interference pattern is recorded on the 16 bits CCD chip
of $1024 \times  1024$ pixels.

The observation of particles in motion imposes to limit the exposition time in order to limit the blurring
of the signal. The magnification calibration of our setup gives a lateral pixel size of 160 nm. In our case, the
diffusion constant of the Brownian motion is equal to $D \simeq 4 \mu \textrm{m}^2.\textrm{s}^{-1}$ according to the Stokes-Enstein law. So
the exposition time must be lower than 3 ms to avoid that the signal is shiffted by one pixel. The exposition
time of our camera can be set under this limit, but the acquisition period during 2 frames is limited to 80 ms.
The travel of 100 nm particles in water during this period is too long to use the phase-shifting technique in order
to improve the signal-to-noise ratio \cite{zhang1998three,yamaguchi2001image}.

\section{3D reconstruction of the scattered light}

\begin{figure}
\begin{center}
 \includegraphics[width=12 cm]{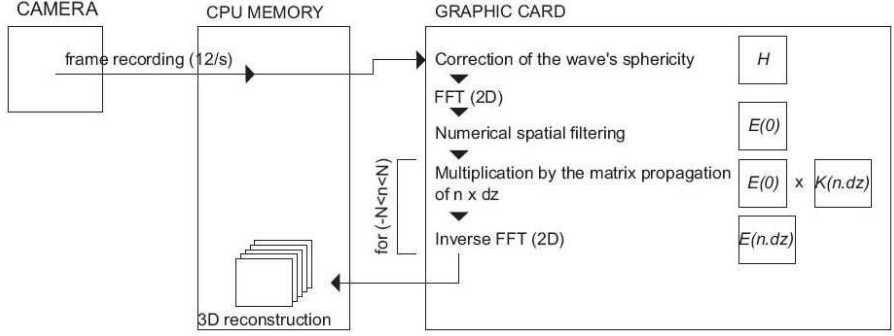}\\
  \caption{Steps of the numerical reconstruction.}\label{Fig2}
\end{center}
\end{figure}

The camera record for each frame an intensity $I_{ccd}$ described in the following equation :
\begin{eqnarray}\label{EQ1}
 I_{ccd} &=& I_{ref} + I_s + E_{ref} · E^*_s + E_s · E^*_{ref}
\end{eqnarray}
where $I_{ref}$ and $I_s$ are respectively the intensity of the reference and the scattering light. $E_{ref}$ and $E_s$ are the
respective fields. The tilted beam splitter which combines the two fields adds a spatial frequency on the cross
terms of interference (the third and fourth terms of the equation (1)). The effect is that these two terms can
be separated in the Fourier space of the holograms. Figure \ref{Fig2} summarizes the steps of our reconstruction. First
we multiply the hologram by a complex phase matrix $
\exp[j\pi (x2 + y2)/(\lambda d)]
$
where $\lambda$ is the wavelength and
$d$ the distance between the CCD and the output pupil of the microscope objective. This operation is similar
to a back-propagation of the hologram to the output plane of the objective. This propagation compensates
the sphericity of the wave added by the objective. Then we calculate the Fast-Fourier Transform (FFT) of the
corrected hologram. In the Fourier plane, the two first term of equation \ref{EQ1} are in the center, while the cross
terms are respectively centered on the frequency added by the off-axis geometry and its complex conjugate. We
numerically filter the term $E_s · E^*_{ref}$ of equation \ref{EQ1} with a circular mask of radius equal to the apparent radius
of the output pupil. The filtered map of the spatial frequencies ($f[k_x, k_y]$) obtained is multiplied by the following
matrix of propagation :
\begin{eqnarray}\label{EQ2}
K[k_x, k_y, n] = \exp\left(\frac{(n\times \delta z ). j\lambda (k_x^2+k_y^2)}{2\pi}\right)
\end{eqnarray}
where $\delta z$ is the arbitrary step of axial propagation (typically 100 nm) and $n$ an integer. Then we calculate
the inverse FFT. The result $\epsilon (n) = \textrm{FFT}^{-1}[f · K]$ is the reconstruction of the field at the distance $n \times  \delta z$. We
repeat this multiplication by varying the integer n to reconstruct the volume. We get a stack of lateral sections
$\epsilon (n· \delta z)$ spaced by $\delta z$. The positions of the scatterers are determined by the local maximum of the intensity of
the whole reconstructed field.

All this numerical reconstruction is implemented on a graphic card unit. We developp our algoritm with the
Nvidia CUDA architecture to parallelize all the calculation on the 448 cores of a Nvidia Geforce GTX 470 card. All the steps of the reconstructions, except for the FFTs, need multiplication pixels to pixels and not matrix
multiplication. This kind of operation are perfectly adapted for parallelization on graphic cores \cite{samson2011video,shimobaba2008real,ahrenberg2009using,kang2009graphics}.
 The FFT
steps are made with the CUFFT library. We divided the time of reconstruction by a factor 30 compare to the
same reconstruction made only on the CPU (double quad-core Intel Xeon E5520 at 2.27 GHz). We developped
a software for live visualisation of reconstruction using OpenGL library to display the datas calculated on the
graphic card.

\section{Current results}

\begin{figure}
\begin{center}
   \includegraphics[width=10 cm]{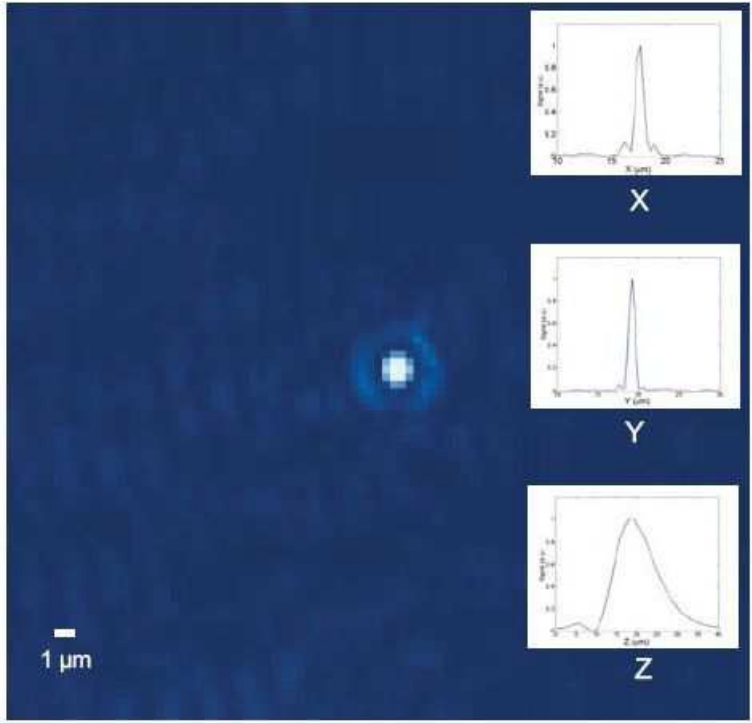}\\
  \caption{ Intensity of the scattered field by a 100 nm gold particle plotted in the section where the intensity is maximum.
The three curves at the left shows the PSF respectively in $X,Y$ and $Z$ directions.}\label{Fig3}
\end{center}
\end{figure}

We are able now to observe 100nm gold nano particles in Brownian motion during few tens of positions. After
that we generally lose the targets. The figure \ref{Fig3} shows the observation we are able to do in the present state of
the setup. We see on the figure the lateral section where the intensity of the scattered light by the nano-bead
is maximum. The PSF of the reconstructed field is plotted for $X, Y$ and $Z$ directions. The lateral PSF width
is equal to 1 $\mu$m, and the axial width is equal to 14 $\mu$m. The signal-to-noise ratio of the reconstruction is better
than 100. These results are obtained with an exposition time of only 1ms and an acquisition period of 80 ms
between two frames.

\section{Conclusion}

We report in this paper the present performance of our digital holographic microscope. It consists of a Mach-Zender setup in off-axis geometry. A dark field microscope objective collects the light scattered by nano-sized
gold particles, which interferes with the reference plane wave on the CCD camera. The scattered light field
is numerically propagated from the interference pattern. We used our own algorithm based on the CUDA
architecture to reduce the calculation time. The improvement clearly shows that this technology is a very
promising tool for digital holography. We are able now to localize 100 nm gold nano particles in Brownian
motion from holograms exposed during 1 ms. We expect to obtain similar performance for gold particles in a
crowed medium or in a living cell in order to study the diffusion law in these specific media.

\bibliographystyle{unsrt}


\end{document}